\documentclass[aps,prd,groupedaddress,showpacs]{revtex4}

\usepackage{graphicx}
\usepackage{dcolumn}
\usepackage{bm}
\usepackage{amsmath}
\usepackage{epstopdf}
\usepackage{amsfonts}
\usepackage{amssymb}
\usepackage{hyperref}
\usepackage{xcolor}

\usepackage[utf8]{inputenc}
\usepackage{natbib}

\providecommand{\U}[1]{\protect\rule{.1in}{.1in}}

\newcommand{\baa}{\begin{align}}
\newcommand{\eaa}{\end{align}}

\newcommand{\be}{\begin{equation}}
\newcommand{\ee}{\end{equation}}
\newcommand{\bea}{\begin{eqnarray}}
\newcommand{\eea}{\end{eqnarray}}

\begin{document}



\title{Quasinormal modes of black holes with a scalar hair
\\
in Einstein-Maxwell-dilaton theory}


\author{\'Angel Rinc\'on}
\affiliation{
Instituto de F\'isica, Pontificia Universidad Cat\'olica de Valpara\'iso,
\mbox{Avenida Brasil 2950, Casilla 4059, Valpara\'iso, Chile.}}
\email{angel.rincon@pucv.cl}

\author{Grigoris Panotopoulos}
\affiliation{Centro de Astrof\'{\i}sica e Gravita{\c c}{\~a}o, Departamento de F{\'i}sica, Instituto Superior T\'ecnico-IST, Universidade de Lisboa-UL, 
Av. Rovisco Pais, 1049-001 Lisboa, Portugal.}
\email{grigorios.panotopoulos@tecnico.ulisboa.pt}

%

\date{\today}

\begin{abstract}
We compute the quasinormal frequencies for scalar perturbations of hairy black holes in four-dimensional Einstein-Maxwell-dilaton theory assuming a non-trivial scalar potential for the dilaton field. We investigate the impact on the spectrum of the angular degree, the overtone number, the charges of the black hole as well as the magnitude of the scalar potential. All modes are found to be stable. Our numerical results are summarized in tables, and for better visualization, we show them graphically as well. 
%
\end{abstract}

\maketitle


\section{Introduction}\label{Intro}

Black holes (BHs), one of the most remarkable predictions of Einstein's
General Relativity (GR) \cite{Einstein:1916vd} and other metric theories of gravity, are
without a doubt fascinating objects of paramount importance both for classical and quantum gravity, linking together several different research areas, from gravitation and astrophysics to quantum mechanics and statistical physics. Their existence has been established over the last years in a two-fold way, namely from the one hand by the first image of a black hole shadow announced last year by the Event Horizon Telescope \cite{L1,L2,L3,L4,L5,L6}
as well as by the numerous direct detection of gravitational waves from BH binary systems by the LIGO/Virgo collaborations 
\cite{ligo1,ligo2,ligo3,ligo4,ligo5}.

\smallskip

The gravitational wave astronomy has opened up a new window to our Universe, and it provides us with an excellent tool to test gravitation under extreme conditions. The signal of the gravitational waves emitted from the merger of the two BHs during the ring down phase is dominated by the fundamental mode of the so called quasinormal modes (QNMs) of the black holes. The QN frequencies $\omega=\omega_R + i \omega_I$ are complex numbers that encode the information on how a black hole relaxes after a perturbation is not applied any longer. Their main features may be summarized as follows: a) They only depend on the space-time and the type of perturbation, and not on the initial conditions. Therefore they are characteristic frequencies of the system, b) They have a non-vanishing imaginary part, $\omega_I$, due to the fact that a perturbed BH responds emitting gravitational waves, and the sign of the imaginary part indicates if the mode is stable or unstable. To be more precise, a mode is unstable (exponential growth) when $\omega_I > 0$, otherwise it is stable (exponential decay) when $\omega_I < 0$, c) in the case where a mode is stable, the real part gives the frequency of the oscillation, $\omega_R/(2 \pi)$, while the inverse of $|\omega_I|$ determines the dumping time, $t_D^{-1}=|\omega_I|$. 
Black hole perturbation theory has been developed thanks to the works of Regge and Wheeler \cite{wheeler}, Zerilli \cite{zerilli1,zerilli2,zerilli3}, Moncrief \cite{moncrief} and Teukolsky \cite{teukolsky}. Chandrasekhar's monograph \cite{monograph} is a standard textbook that contains the mathematics of black holes, while for reviews of the same topic see \cite{review1,review2,review3}.

\smallskip

According to the standard lore, ideal isolated black holes are characterized by a small number of parameters, namely the mass, the rotation speed and the electric or magnetic charges. It is possible, however, to obtain less conventional black hole solutions with a scalar hair if at least one of the conditions that forbid their existence is violated. Those hairy black holes are then characterized by additional properties (scalar charge), which is not associated to a Gauss' law, and which may or may be not an independent quantity. In the latter case it is called secondary charge, and it will be given in terms of the other properties of the black hole, such as mass and electric charge. For an excellent review see e.g. \cite{CarlosReview}.

\smallskip

Given the rapid development of the field and the interest of the community in QNMs of BHs, 
the spectra expected from less conventional BH solutions should be investigated as well. 
In the present article we propose to compute the quasinormal frequencies of hairy black holes in four-dimensional Einstein-Maxwell-dilaton (EMD) theory inspired by Supergravity \cite{Nilles} and also by the low-energy effective action of Superstring Theories \cite{ST1,St1b,ST2,ST2b}. The properties of black holes solutions obtained in linear EMD \cite{Clement:2002mb} have been studied thoroughly in the literature \cite{Clement:2007tw,Bertoldi:2009yi,Pasaoglu:2009te,Sakalli:2010yy,Sakalli:2014wja,Sakalli:2016abx,
Sakalli:2016fif,Sakalli:2016jkf}. To the best of our knowledge, there is currently only one work on QNMs of BHs with a scalar hair, albeit in a different theory \cite{Chowdhury:2018izv}. 

\smallskip

The plan of our work is organized as follows: After this Introduction, we briefly review, in the next section, the gravitational background as well as the wave equation with the corresponding effective potential barrier for scalar perturbations. In section 3 we compute the quasinormal frequencies employing the WKB approximation of 6th order, and we discuss our results. Finally, in the fourth section we summarize our work with some concluding remarks.

\section{Gravitational background and wave equation} 
\label{Classical}
\noindent

\subsection{The model}

We start by introducing the model (see \cite{Astefanesei:2019qsg} and references therein), which consists of the usual Einstein-Hilbert term for gravity plus the Maxwell-dilaton Lagrangian for matter, where a concrete dilaton potential is considered. The model is described by the following action
\begin{align}
\begin{split}
S[g_{\mu \nu}, A_{\mu},\phi] \equiv \frac{1}{2\kappa}\int \mathrm{d}^4 x \sqrt{-g} 
\bigg[
&
R - \text{e}^{\gamma \phi} F_{\mu \nu} F^{\mu \nu} - 
\frac{1}{2}\partial_{\mu}\phi \partial^{\mu}\phi - V(\phi)
\bigg]
\end{split}
\end{align}
where $\kappa \equiv 8 \pi G$ with $G$ being Newton's constant, $g$ is the determinant of the metric tensor $g_{\mu \nu}$, $R$ is the corresponding Ricci scalar, $F_{\mu \nu} \equiv \partial_\mu A_\nu - \partial_\nu A_\mu $ is the electromagnetic field strength, and $\phi$ is the dilaton field. The latter is non-minimally coupled to the electromagnetic Lagrangian with a coupling constant $\gamma$. In the discussion to follow, we shall use geometrical units such as $\kappa=1=c$. Finally, for the dilaton field we shall consider a non-trivial potential given by
\begin{align}
V(\phi) &= 2 \alpha [2 \phi + \phi \cosh(\phi) - 3 \sinh(\phi)]
\end{align}
with $\alpha$ being a dimensionfull parameter with dimensions $[\alpha] = L^{-2}$.

As was pointed out in \cite{Anabalon:2013qua} and subsequently in \cite{Astefanesei:2019qsg}, the scalar potential assumed here was originally introduced to obtain exact solutions in EMD gravity. Moreover, the theory described by the action $S[g_{\mu \nu}, A_{\mu}, \phi]$ can also be considered as a consistent truncation of $\mathcal{N} = 2$ supergravity in four space-time dimensions, coupled to a vector multiplet and deformed by a Fayet-Iliopoulos term \cite{fayet}, see e.g.~ \cite{Anabalon:2017yhv,Astefanesei:2019qsg} for further details.

The field equations for the metric tensor, the Maxwell potential and the dilaton are found to be \cite{Astefanesei:2019ehu,Astefanesei:2019qsg}
\begin{align}
G_{\mu \nu} &= \frac{1}{2} \Bigl( T_{\mu \nu}^{(\phi)} + T_{\mu \nu}^{(M)} \Bigl)
\\
\partial_{\mu} \Bigl( \sqrt{-g}\text{e}^{\gamma \phi} F^{\mu \nu} \Bigl) &= 0
\\
\frac{1}{\sqrt{-g}}\partial_{\mu} \Bigl( \sqrt{-g} g^{\mu \nu} \partial_{\nu} \phi \Bigl) &= \frac{\mathrm{d}V(\phi)}{\mathrm{d}\phi} + \gamma \text{e}^{\gamma \phi} F^2
\end{align}
respectively, where the energy-momentum tensor associated to the dilaton and the Maxwell potential are computed to be \cite{Astefanesei:2019ehu,Astefanesei:2019qsg}
\begin{align}
T_{\mu \nu}^{(\phi)} &\equiv \partial_{\mu} \phi \partial_{\nu} \phi - g_{\mu \nu} \bigg[ \frac{1}{2} (\partial \phi)^2  + V(\phi) \bigg]
\\
T_{\mu \nu}^{(M)} &\equiv 4 \text{e}^{\gamma \phi} \bigg(F_{\mu \beta}F^{\beta}_{\nu} - \frac{1}{4}g_{\mu \nu} F^2 \bigg)
\end{align}
respectively.

\subsection{Black hole solution}

For static, spherically symmetric solutions in Schwarzschild coordinates, $(t,r,\theta,\varphi)$, and adopting the mostly positive metric signature $(-,+,+,+)$, we make as usual the following ansatz for the line element:
\begin{align}
\mathrm{d}s^2 &= -f(r) \mathrm{d}t^2 + g(r)^{-1}\mathrm{d}r^2 + r^2 \mathrm{d}\Omega_2^2
\end{align}
where $\mathrm{d}\Omega_2^2 \equiv \mathrm{d}\theta^2 + \sin^2(\theta)\mathrm{d}\varphi^2$, 
while $f(r)$ and $g(r)$ are two unknown functions of the radial coordinate. Equivalently, one may 
introduce two new functions $\sigma(r),m(r)$ as follows
\begin{align}
f(r) &\equiv g(r)\sigma(r)^2
\\
g(r) &\equiv 1 - \frac{2 m(r)}{r}
\end{align}
with $m(r)$ being the Misner-Sharp mass function \cite{Misner:1964je}, while the red-shift 
function $\sigma(r)$ is computed to be \cite{Astefanesei:2019qsg}
\begin{align}
\sigma(r) &= \left(1 + \frac{Q_s^2}{4r^2} \right)^{-1/2} 
\end{align}
where $Q_s$ is the scalar charge of the hairy black hole. Accordingly, the  Misner-Sharp mass function is found to be \cite{Astefanesei:2019qsg}
\begin{align}
\begin{split}
m(r) = \frac{r}{2 \sigma^2(r)} 
&\bigg\{
\frac{Q_e^2}{r^2} \bigg[\chi(r)-1 \bigg] - 
\alpha
\bigg[ 
\frac{Q_s^2}{2}\chi(r) - r^2 \phi(r)
\bigg]
\bigg\}
-
\frac{Q_s^2}{8r}
\end{split}
\end{align}
where $Q_e$ is the electric charge of the black hole.
The auxiliary function $\chi$ is defined by
\begin{align}
\chi(r) \equiv \sqrt{1 + \frac{4r^2}{Q_s^2}}
\end{align}
Notice that the black hole solution with a scalar hair obtained in the model considered here is characterized by two charges $\{Q_e, Q_s\}$. The electric charge is associated to a Gauss' law, and it can be formally defined in the following way \cite{Astefanesei:2019qsg}
\begin{align}
Q_e &\equiv \frac{1}{8 \pi} \oint_S \mathrm{d}S_{\mu \nu} \text{e}^{\gamma \phi} F^{\mu \nu}
\end{align}
In absence of sources to Maxwell's equations, the corresponding charge is independent of the surface $S$.
What is more, we can also identify the ADM mass taking advantage of the asymptotic value of $m(r)$, 
namely \cite{Astefanesei:2019qsg}:
\begin{align}
M &= \frac{Q_e^2}{Q_s} -\frac{1}{12}\alpha Q_s^3 
\end{align}
Therefore the scalar charge may be computed in terms of the mass and the electric charge. This implies that $Q_s$ is a secondary charge, according to the terminology of \cite{CarlosReview}, since it is not an independent quantity.

\section{Scalar perturbations}

We shall now perturb the black hole background, presented in the previous subsection, with a probe scalar field, and we study its propagation in a fixed gravitational space-time. A massless canonical scalar field $\Phi$ is described by a Lagrangian density of the form
\begin{align}
\mathcal{L} &= \frac{1}{2} \Bigl( \partial \Phi \Bigl)^2 
\end{align}
and the corresponding Klein-Gordon equation is given by
\begin{align}
\frac{1}{\sqrt{-g}} \partial_{\mu} 
\Bigl(
\sqrt{-g} g^{\mu \nu} \partial_{\nu}
\Bigl) \Phi &= 0
\end{align}
In a four-dimensional space-time, and in the Schwarzschild coordinates used here, we attempt to find solutions for the scalar field applying the method of separation of variables, making as usual the following ansatz:
\begin{align}
\Phi (t,r,\theta,\varphi) &= \text{e}^{-i\omega t}\frac{\Psi(r)}{r}Y_l^m(\theta, \varphi)
\end{align}
where $Y_l^m(\theta, \varphi)$ are the spherical harmonics, while $\omega$ is the frequency to be computed. 
It is not difficult to show that the radial part satisfies an ordinary second order linear differential equation,
and after a straightforward calculation one obtains for the unknown function $\Psi$ a Schr{\"o}dinger-like equation
\begin{equation}
\frac{\mathrm{d}^2 \Psi}{\mathrm{d}x^2} + [ \omega^2 - V(x) ] \Psi = 0
\end{equation}
with $x$ being the so called tortoise coordinate, which is computed by
\begin{equation}
x  \equiv  \int \frac{\mathrm{d}r}{\sqrt{g(r)f(r)}}
\end{equation}
while the corresponding effective potential barrier is found to be \cite{Konoplya:2006rv,Zinhailo:2018ska}
\begin{widetext}
\begin{equation}
V_s(r) = \frac{l (l+1) f(r)}{r^2} + \frac{f'(r) g(r) + f(r) g'(r)}{2r}
\end{equation}
\end{widetext}
where the prime denotes differentiation with respect to $r$, while $l \geq 0$ is the angular degree.

\begin{figure*}[ht]
\centering
\includegraphics[width=0.32\textwidth]{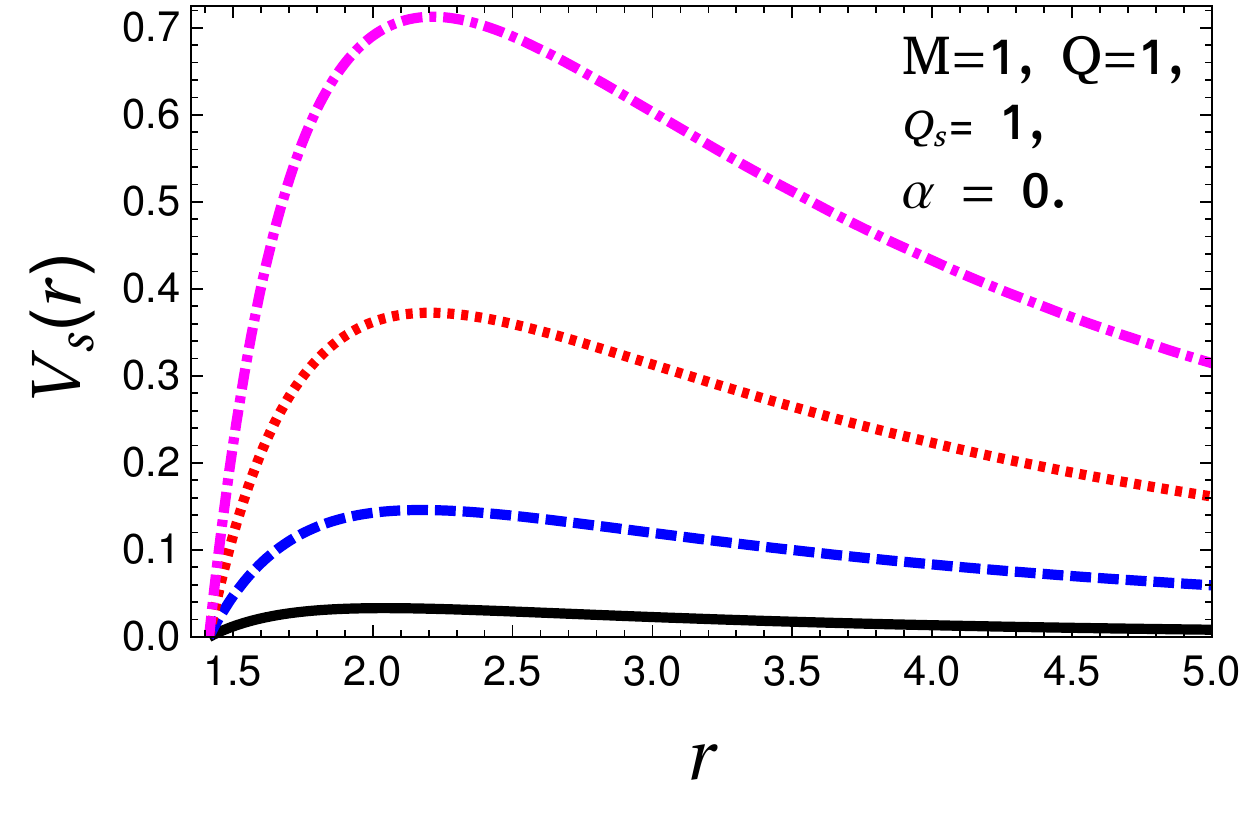}   \
\includegraphics[width=0.32\textwidth]{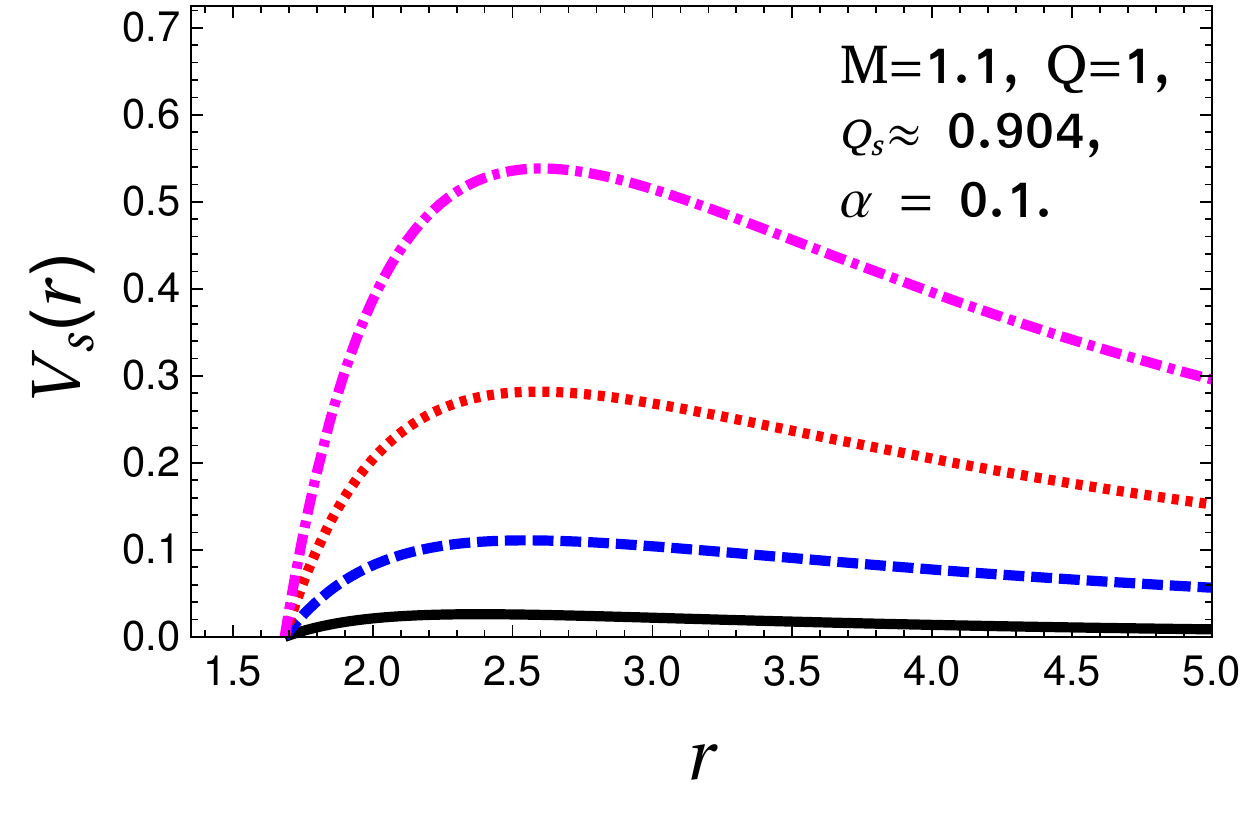}   \
\includegraphics[width=0.32\textwidth]{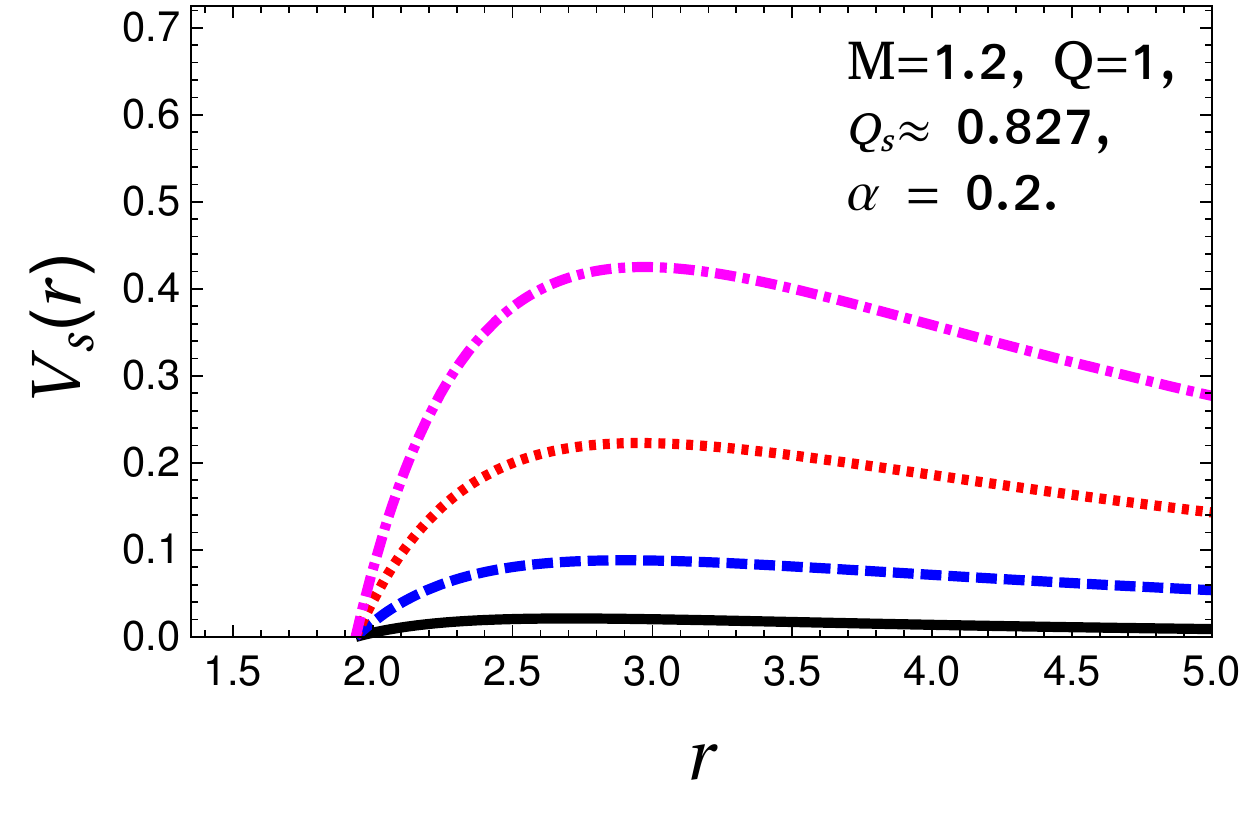}   
\\
\includegraphics[width=0.32\textwidth]{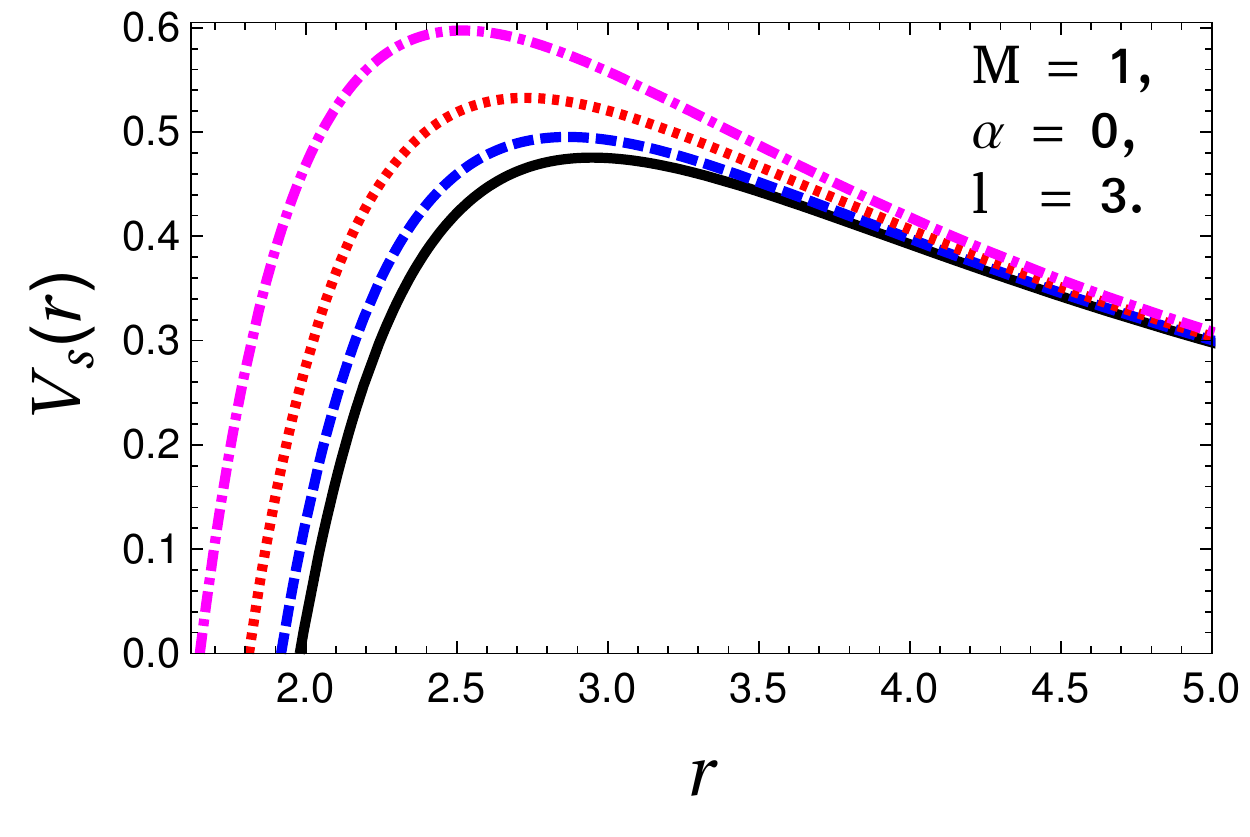}   \
\includegraphics[width=0.32\textwidth]{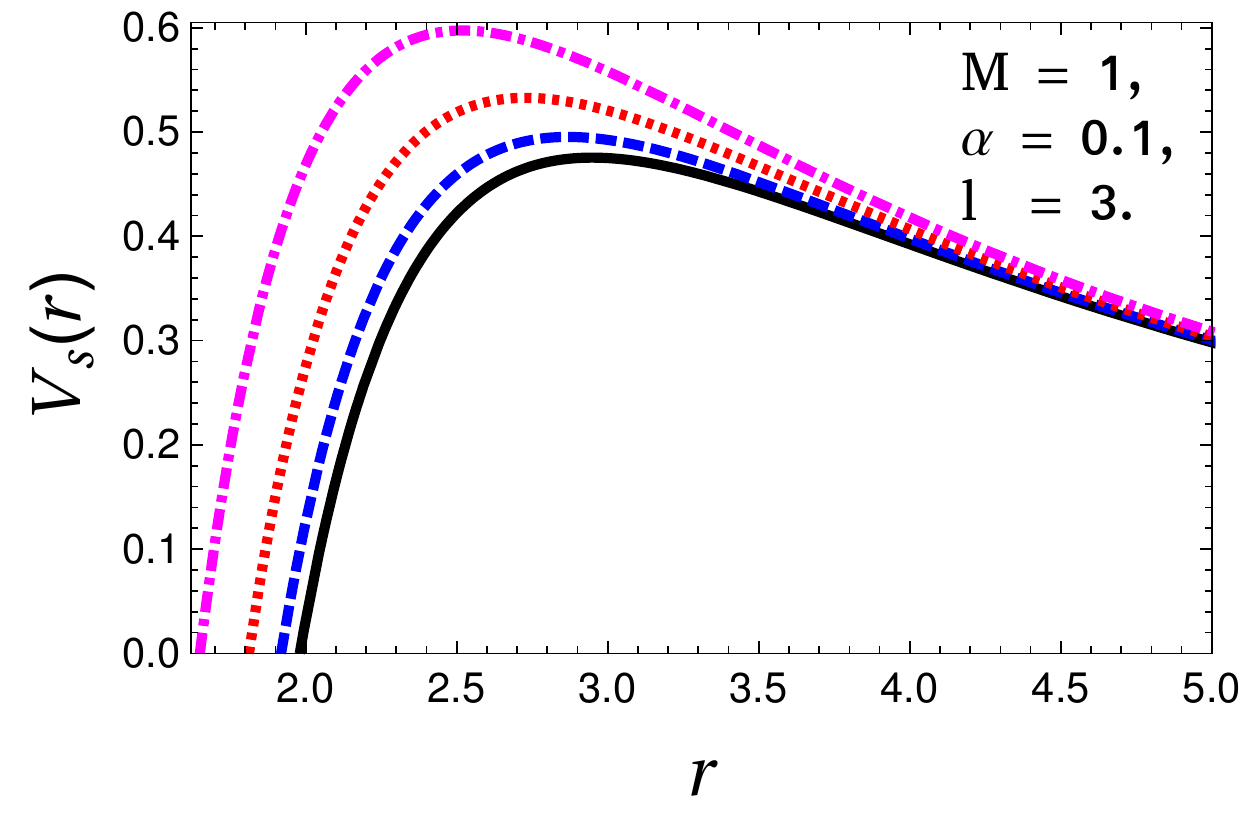}   \
\includegraphics[width=0.32\textwidth]{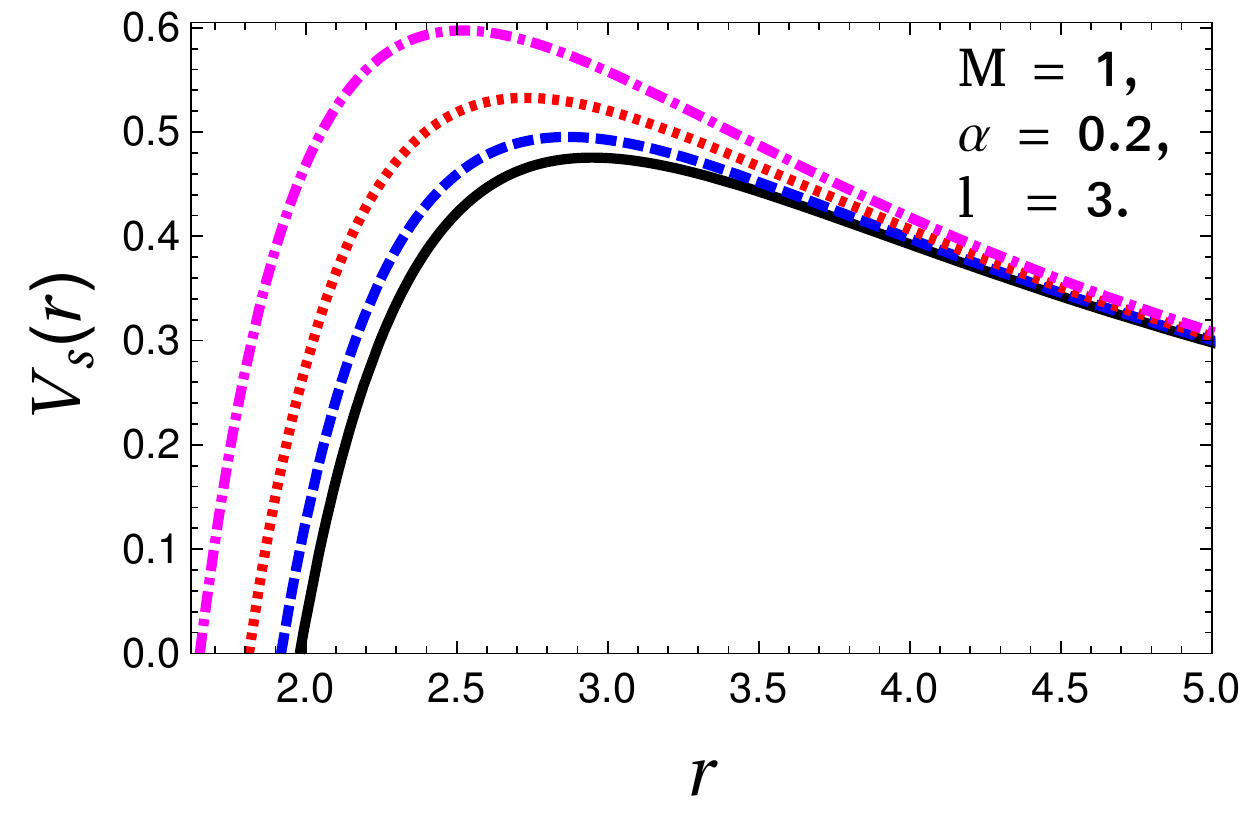}   
\caption{
Effective potential barrier for scalar perturbations, $V_s(r)$, as a function of the radial coordinate $r$ assuming three different values of the parameter $\alpha$.
The panels in the first (left), second (middle) and third (right) show $V_{s}(r)$ for: i) $\alpha = 0$ (case 1), ii) $\alpha=0.1$ (case 2) and iii) $\alpha=0.2$ (case 3), respectively. 
In the first row we have taken $M, Q$, and $Q_s$ as fixed values. Also, they are shown in each sub-figure. Thus, the color code for the first row is given as follow: 
i) $l = 0$ (solid black line), 
ii) $l = 1$ (dashed blue line),
iii) $l = 2$ (dotted red line).
iv) $l=3$ (dot-dashed purple line).
In the second row we have taken $M$ and $l$ as fixed values. Also, can be observed in the corresponding sub-figure the numerical values used. The color code for the second row is then given as follow: 
{\bf LEFT:}
i) $Q=0.2$ and $Q_s= 0.04$ (solid black line), 
ii) $Q=0.4$ and $Q_s= 0.16$  (dashed blue line),
iii) $Q=0.6$ and $Q_s= 0.32$  (dotted red line).
iv) $Q=0.8$ and $Q_s= 0.64$  (dot-dashed purple line).
{\bf MIDDLE:}
i) $Q=0.2$ and $Q_s= 0.04$ (solid black line), 
ii) $Q=0.4$ and $Q_s= 0.159994$  (dashed blue line),
iii) $Q=0.6$ and $Q_s= 0.359860$  (dotted red line).
iv) $Q=0.8$ and $Q_s= 0.638614$  (dot-dashed purple line).
{\bf RIGHT:}
i) $Q=0.2$ and $Q_s= 0.04$ (solid black line), 
ii) $Q=0.4$ and $Q_s= 0.159989$  (dashed blue line),
iii) $Q=0.6$ and $Q_s= 0.359721$  (dotted red line).
iv) $Q=0.8$ and $Q_s= 0.637251$  (dot-dashed purple line).
}
\label{fig:potential}
\end{figure*}

Notice that since in the hairy black hole solution studied in this work the two metric potentials $f(r),g(r)$ are different, the effective potential barrier is found to be slightly more complicated in comparison with the usual case where $g(r)=f(r)$. It is easy to verify that in the special case in which $\sigma(r) = 1$, our expressions boil down to the standard ones, namely
\begin{align}
f(r) & = g(r) 
\\
x & = \int \frac{\mathrm{d}r}{f(r)}
\\
V_s(r) & = f(r) \: \left(\frac{l (l+1)}{r^2}+\frac{f'(r)}{r} \right) 
\end{align}
The effective potential barrier as a function of the radial coordinate is shown in \eqref{fig:potential} for several different values of the angular degree $l$, the electric charge of the black hole $Q_e$ and the magnitude of the scalar potential $\alpha$ . 
Left, middle and right panels of first row show $V_s(r)$ for $\alpha = 0$, $\alpha =0.1$ and $\alpha =0.2$ respectively. See each row for specific details.
%
%
Clearly, the effective potential barrier increases both with $Q_e$ and $l$, whereas it is not sensitive to the variation of the $\alpha$ parameter, as it can be seen in the second raw of Fig.~\ref{fig:potential}. In the first raw the potential seems to decrease with $\alpha$, but this is misleading since other parameters also vary at the same time. 

Finally, to complete the formulation of the physical problem we must also impose the appropriate boundary conditions, both at infinity (no radiation is incoming from infinity) and at the horizon (nothing can escape from the horizon of the BH). For asymptotically flat space-times the appropriate boundary conditions to be imposed are given by \cite{valeria}
\begin{equation} \label{arbitrary}
\Psi(x) \rightarrow
\left\{
\begin{array}{lcl}
A e^{-i \omega x} & \mbox{ if } & x \rightarrow - \infty \\
&
&
\\
C e^{i \omega x}  & \mbox{ if } & x \rightarrow + \infty
\end{array}
\right.
\end{equation}
where $A,C$ are two arbitrary coefficients.


\section{QN frequencies of hairy black holes}

As far as QNMs computations are concerned, exact analytic expressions may be obtained in a few cases, see for instance \cite{Cardoso:2001hn,Birmingham:2001hc,Poschl:1933zz,
ferrari,Fernando:2003ai,Fernando:2008hb,Rincon:2018ktz,Destounis:2018utr}.
In most of the cases, however, in order to compute the QN spectra one is obliged to develop or adopt one of the currently available numerical methods. Among the approaches used to achieve that, one method is ``Evolving the time dependent wave equation'' \cite{Vishveshwara:1970zz}, which has some advantages, due to the fact that we do not need to be extremely careful about the adequate boundary conditions on the horizon or at infinity. A second approach called ``Integration of the Time Independent Wave Equation'' was first used by Chandrasekhar and Detweiler \cite{Chandrasekhar:1975zza}. The key point in this method relies on the assumption that a QNM is a solution of an ``incoming waves on the horizon and outgoing at infinity'' \cite{Kokkotas:1999bd}. Thus, we can take an expansion of the Zerilli wave equation \cite{Andersson:1995wu,Andersson:1998ze,Andersson:1998qs}
at horizon and infinity of the form given by \eqref{arbitrary}. They found initial values for the numerical integration of the equation \cite{Kokkotas:1999bd}.
In addition to those, there is also the popular and extensively used semi-analytic WKB method \cite{wkb1,wkb2,wkb3}, and recently\cite{Konoplya:2019hlu}. For an incomplete list see e.g. \cite{paper1,paper2,paper3,paper4,paper5,paper6,Flachi:2012nv}, and for more recent works \cite{paper7,paper8,paper9,paper10,Rincon:2018sgd,
Panotopoulos:2017hns,Panotopoulos:2019qjk,
Panotopoulos:2019gtn,Panotopoulos:2020,
Cardoso:2017soq,Konoplya:2019hml,Oliveira:2018oha,Rincon:2020iwy}, and references therein.

Within the WKB approximation the QN frequencies are given by
\begin{equation}
\omega^2 = V_0+(-2V_0'')^{1/2} \Lambda(n) - i \nu (-2V_0'')^{1/2} [1+\Omega(n)]
\end{equation}
where $n=0,1,2...$ is the overtone number, $\nu=n+1/2$, $V_0$ is the maximum of the effective potential, $V_0''$ is the second derivative of the effective potential evaluated at the maximum, while $\Lambda(n), \Omega(n)$ are complicated expressions of $\nu$ and higher derivatives of the potential evaluated at the maximum. The interested reader may consult for instance \cite{paper2,paper7} where the expressions can be found. In the present work we have used the Wolfram Mathematica \cite{wolfram} code with WKB of 6th order presented in \cite{code}, and we have considered the cases $n \leq l$ only, since it is known in the literature that the WKB approach works very well up to that level, see e.g. tables II, III, IV and V of \cite{TablesRef}.


\begin{figure*}[ht]
\centering
\includegraphics[width=0.32\textwidth]{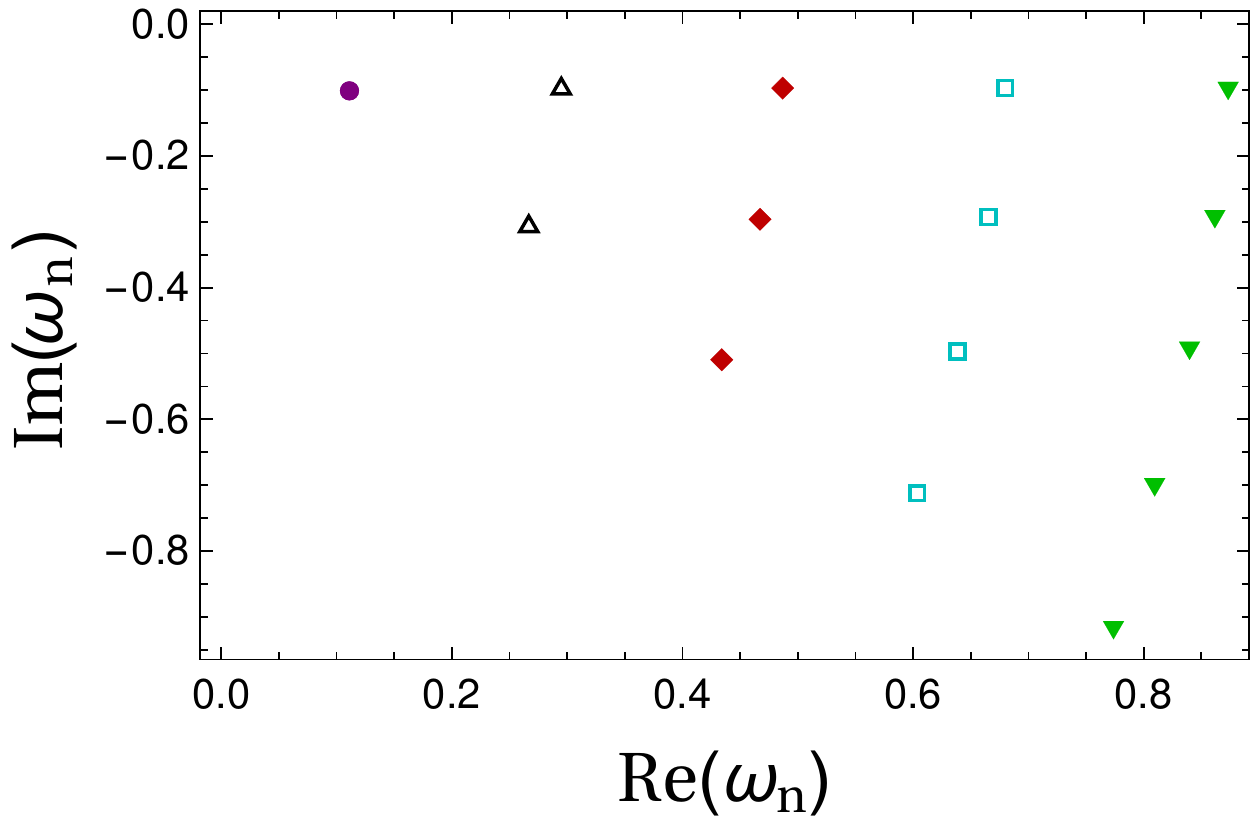} \   
\includegraphics[width=0.32\textwidth]{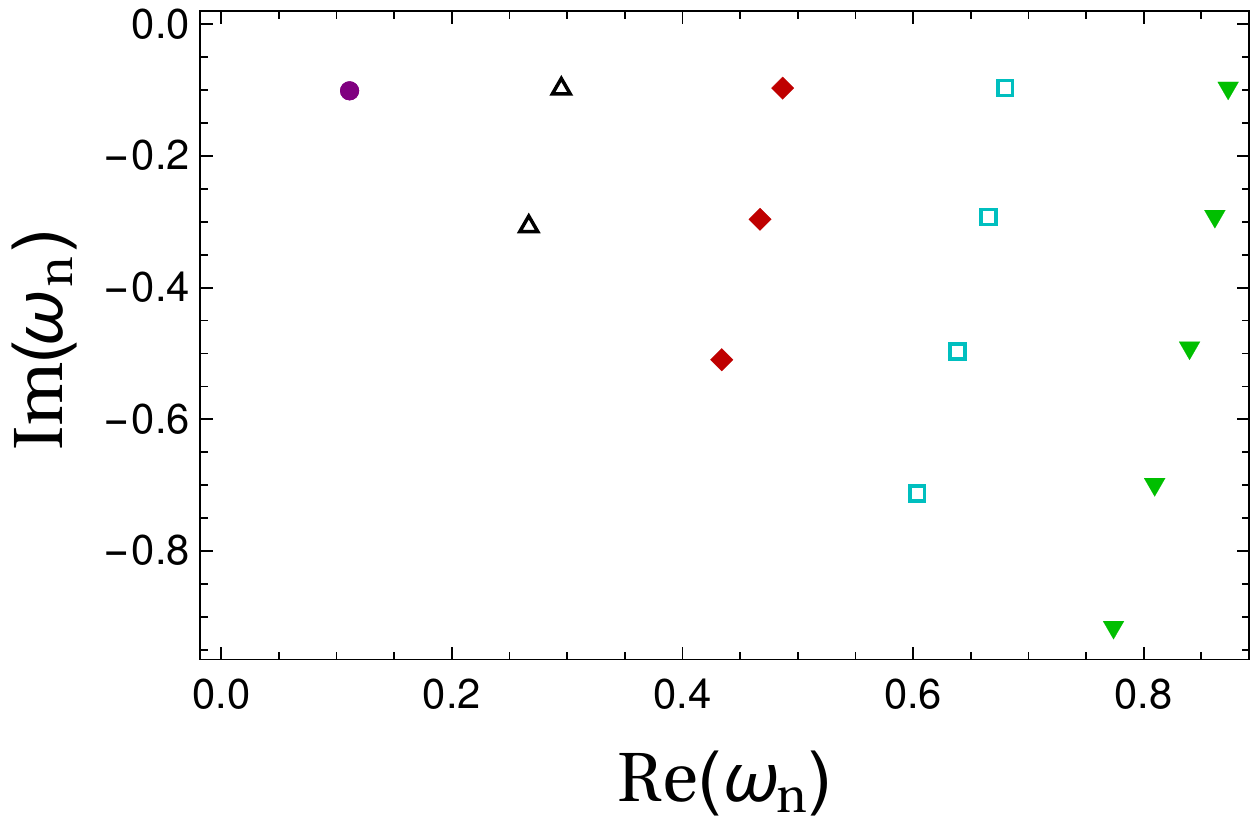} \  
\includegraphics[width=0.32\textwidth]{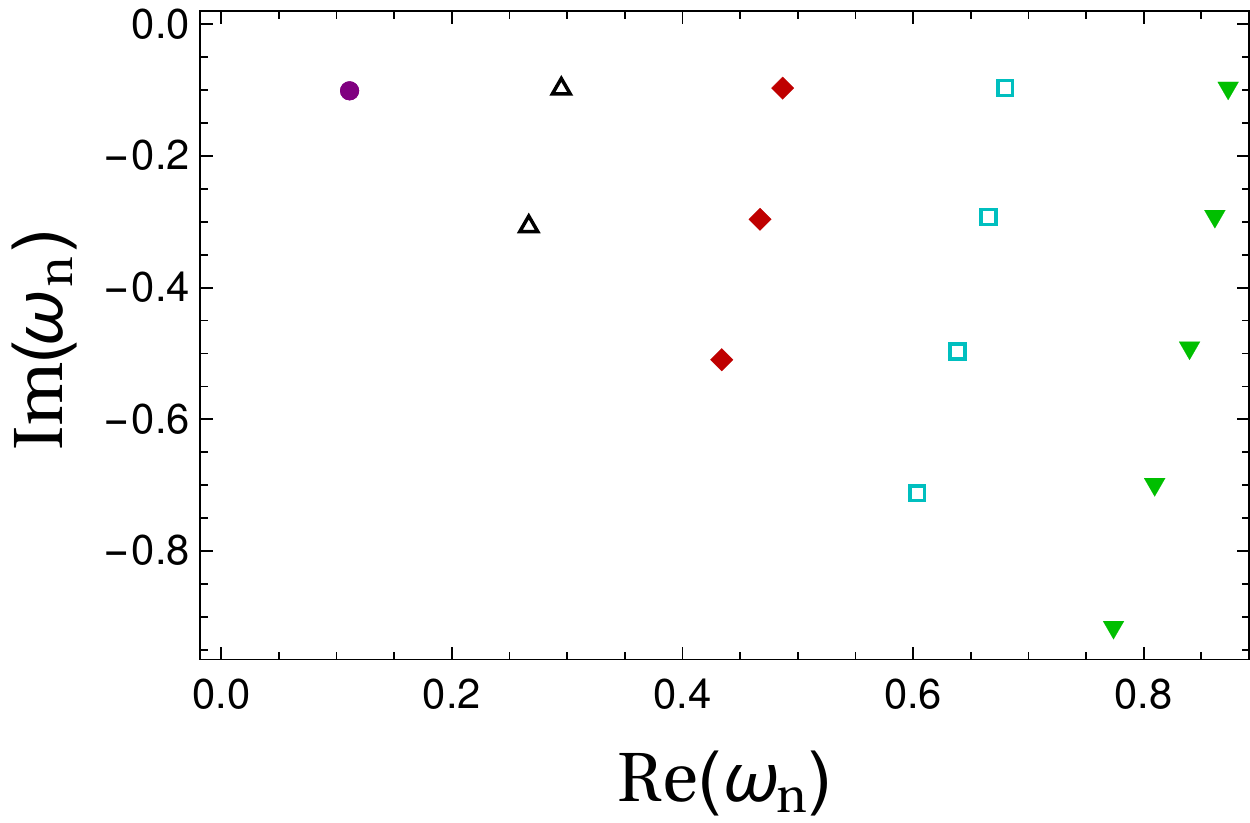}
\\ 
\includegraphics[width=0.32\textwidth]{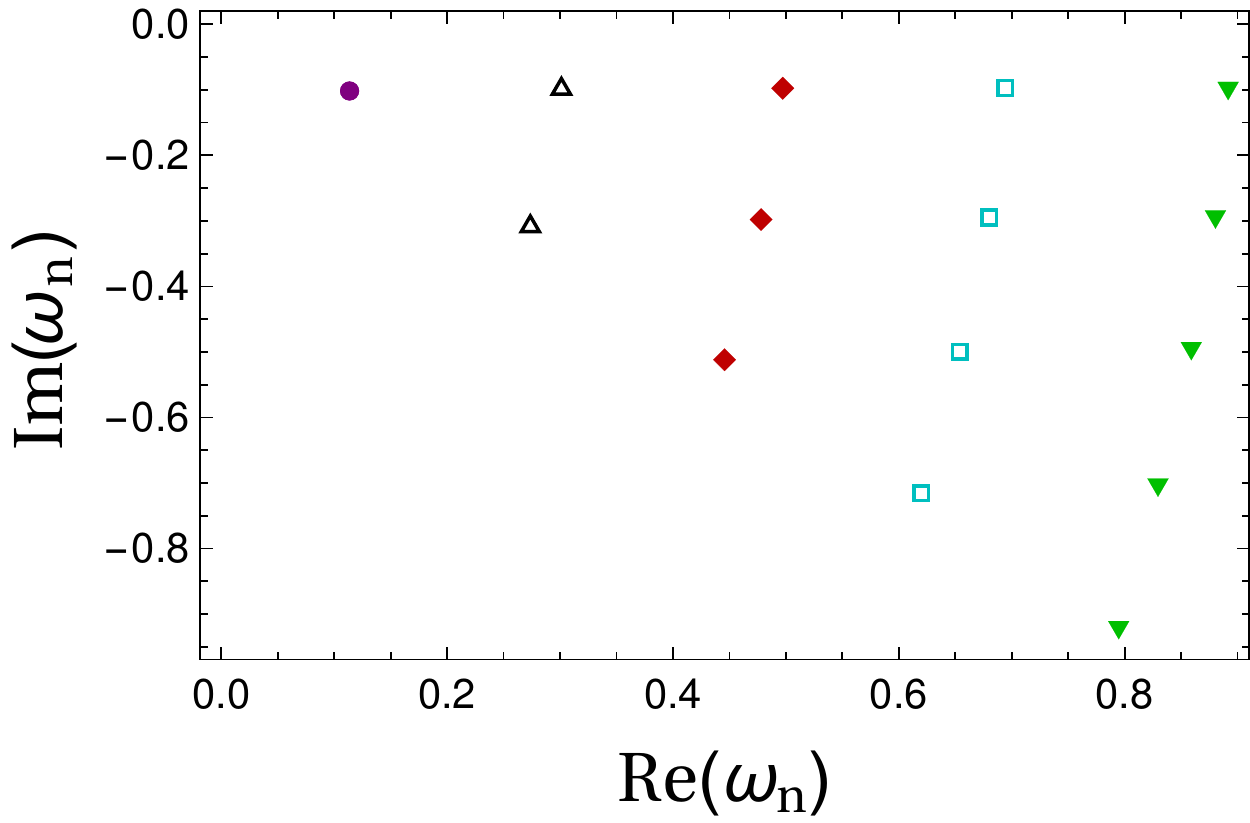} \   
\includegraphics[width=0.32\textwidth]{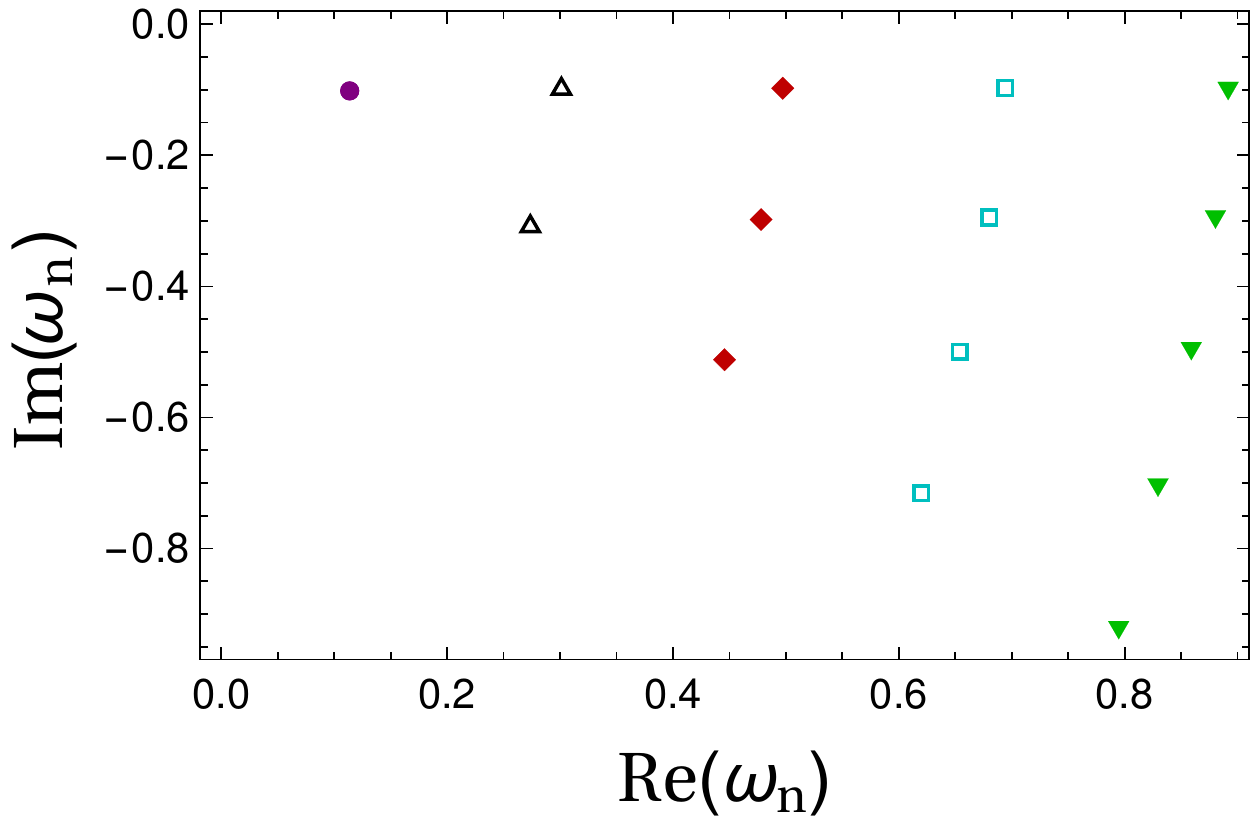} \  
\includegraphics[width=0.32\textwidth]{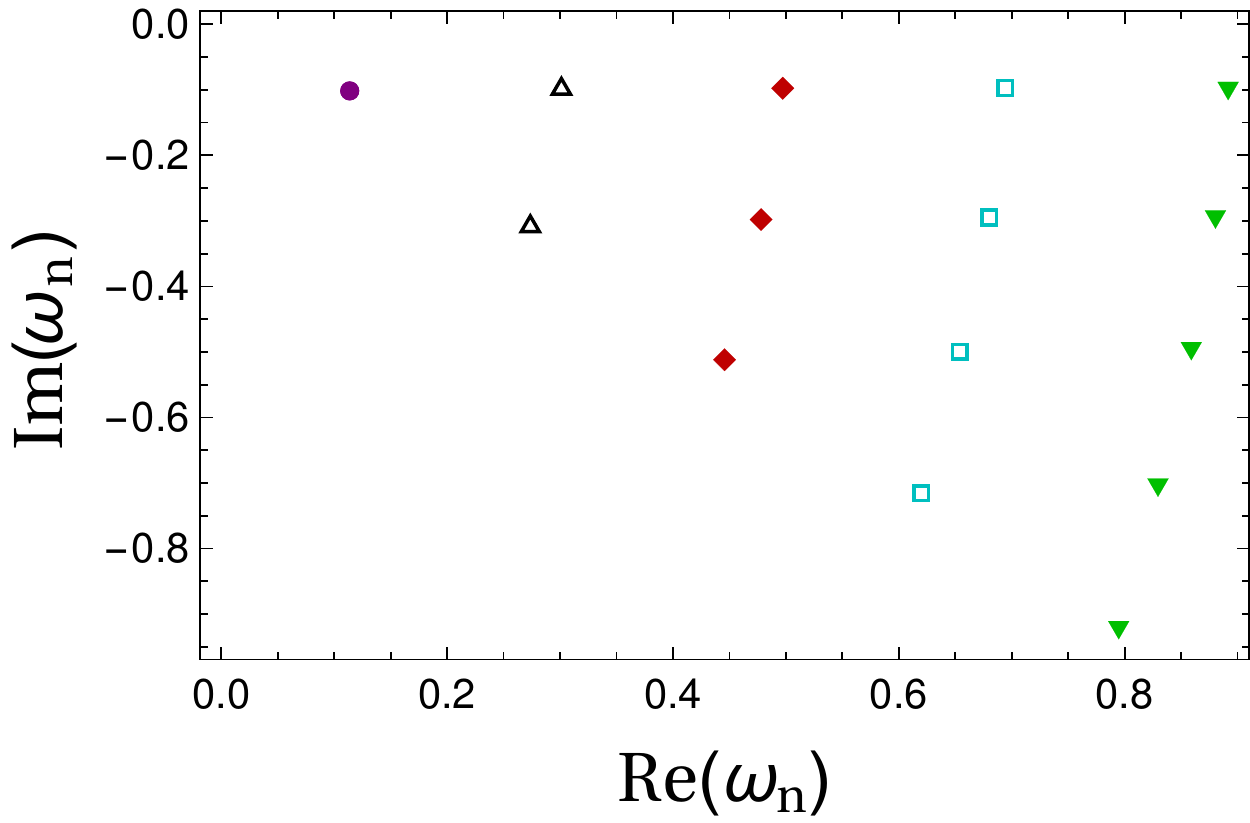}
\\ 
\includegraphics[width=0.32\textwidth]{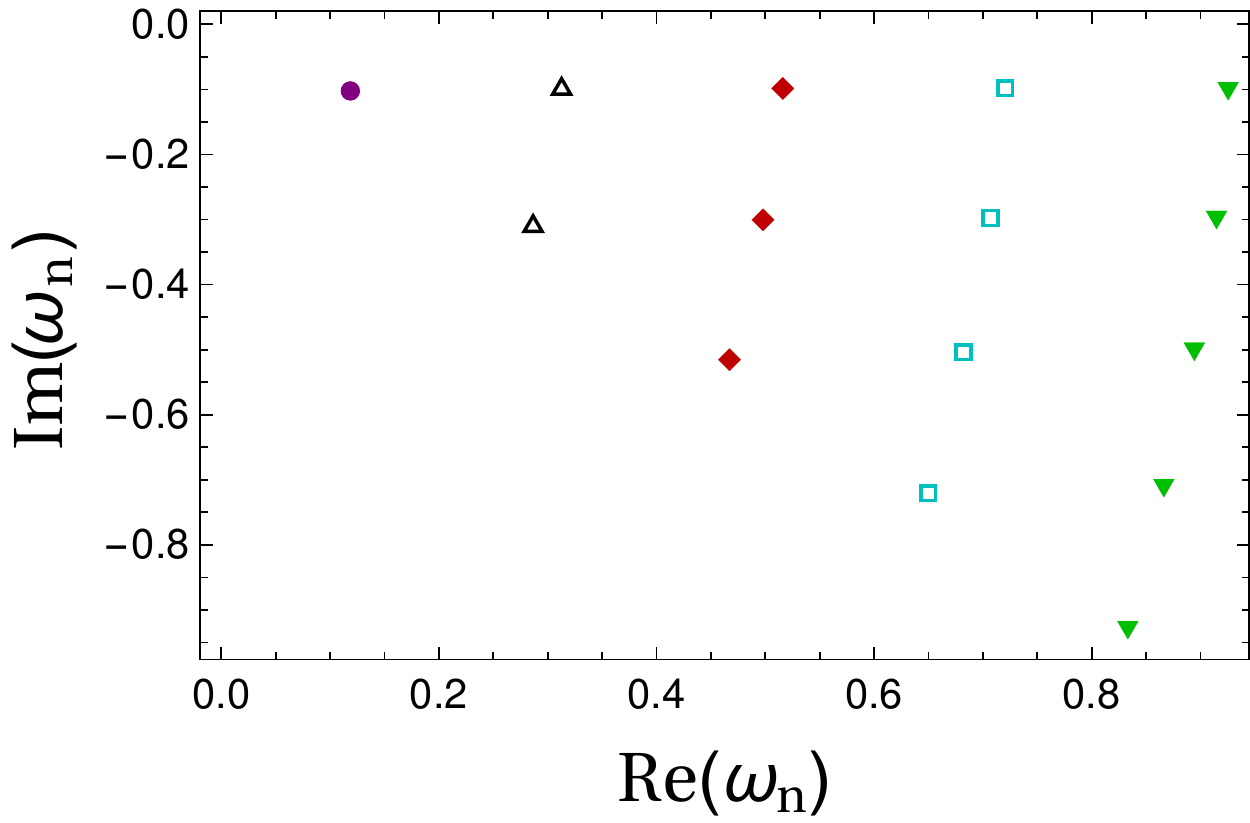} \   
\includegraphics[width=0.32\textwidth]{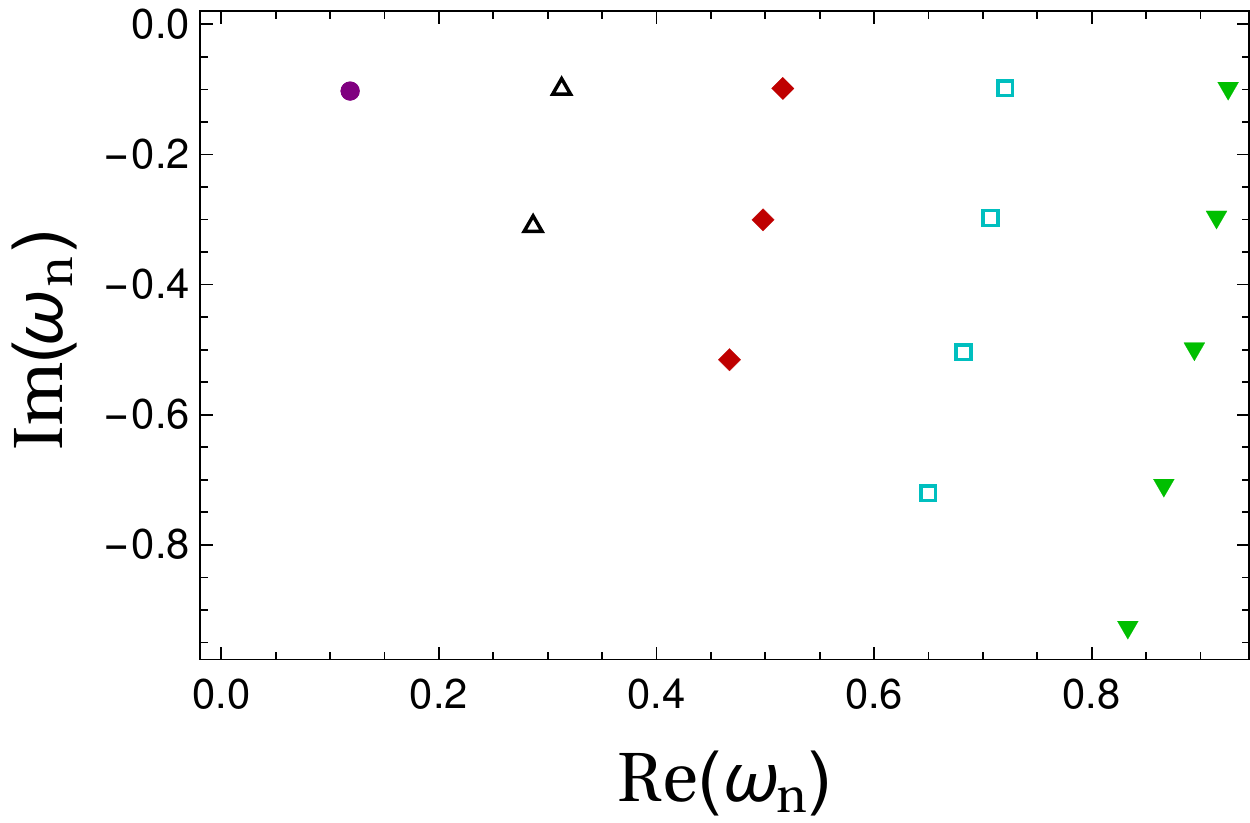} \  
\includegraphics[width=0.32\textwidth]{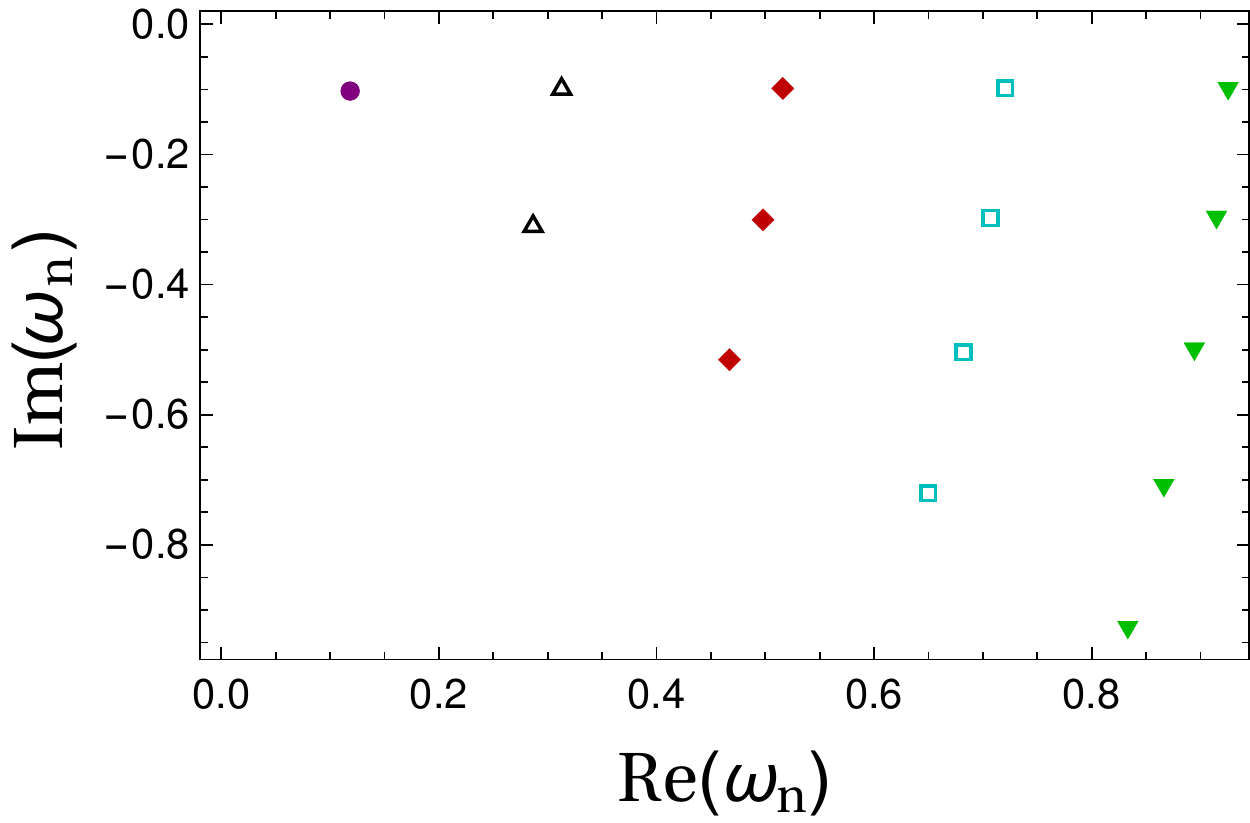}
\\ 
\includegraphics[width=0.32\textwidth]{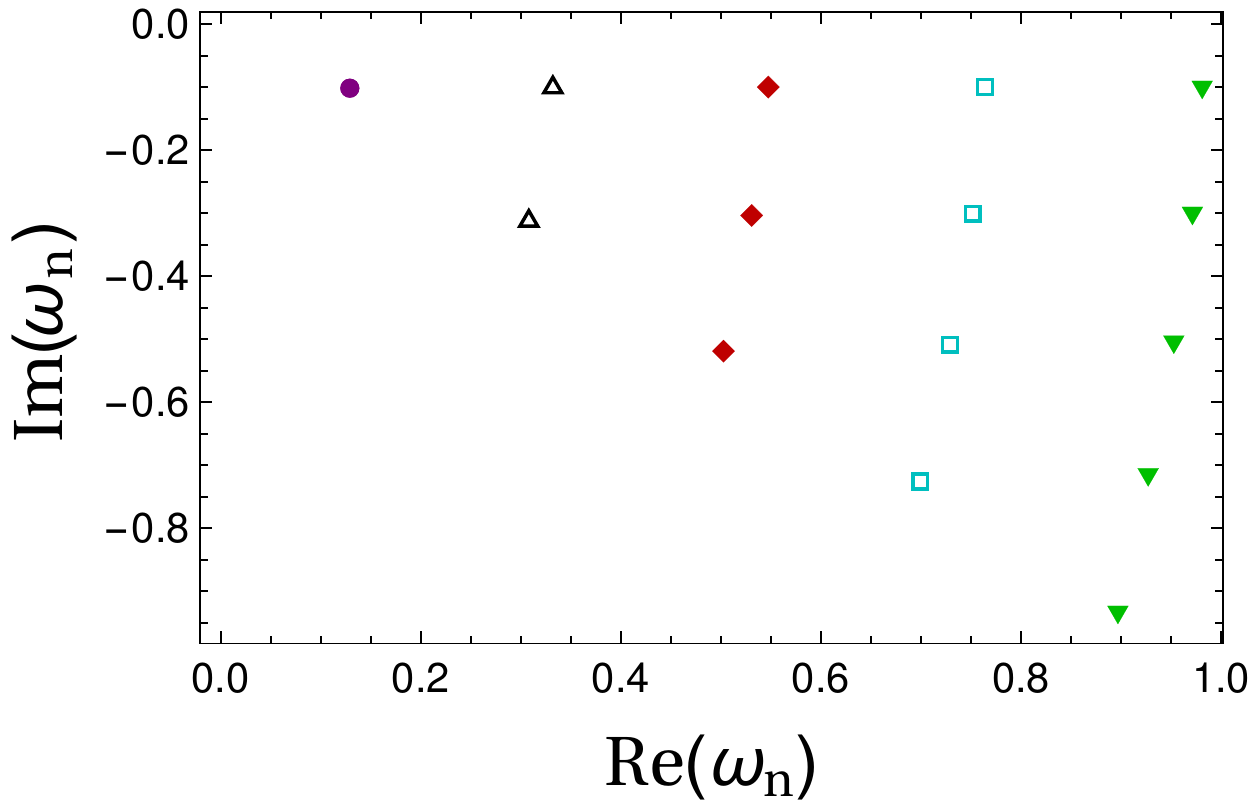} \   
\includegraphics[width=0.32\textwidth]{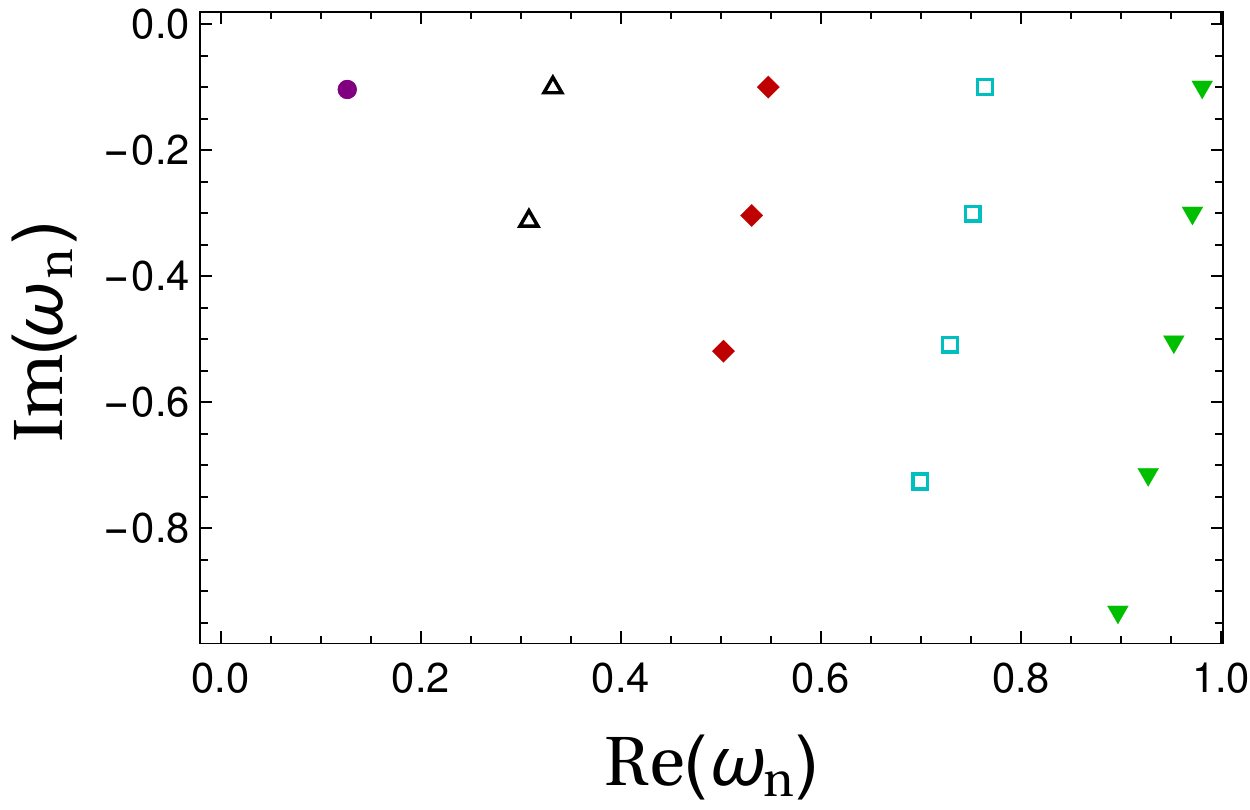} \  
\includegraphics[width=0.32\textwidth]{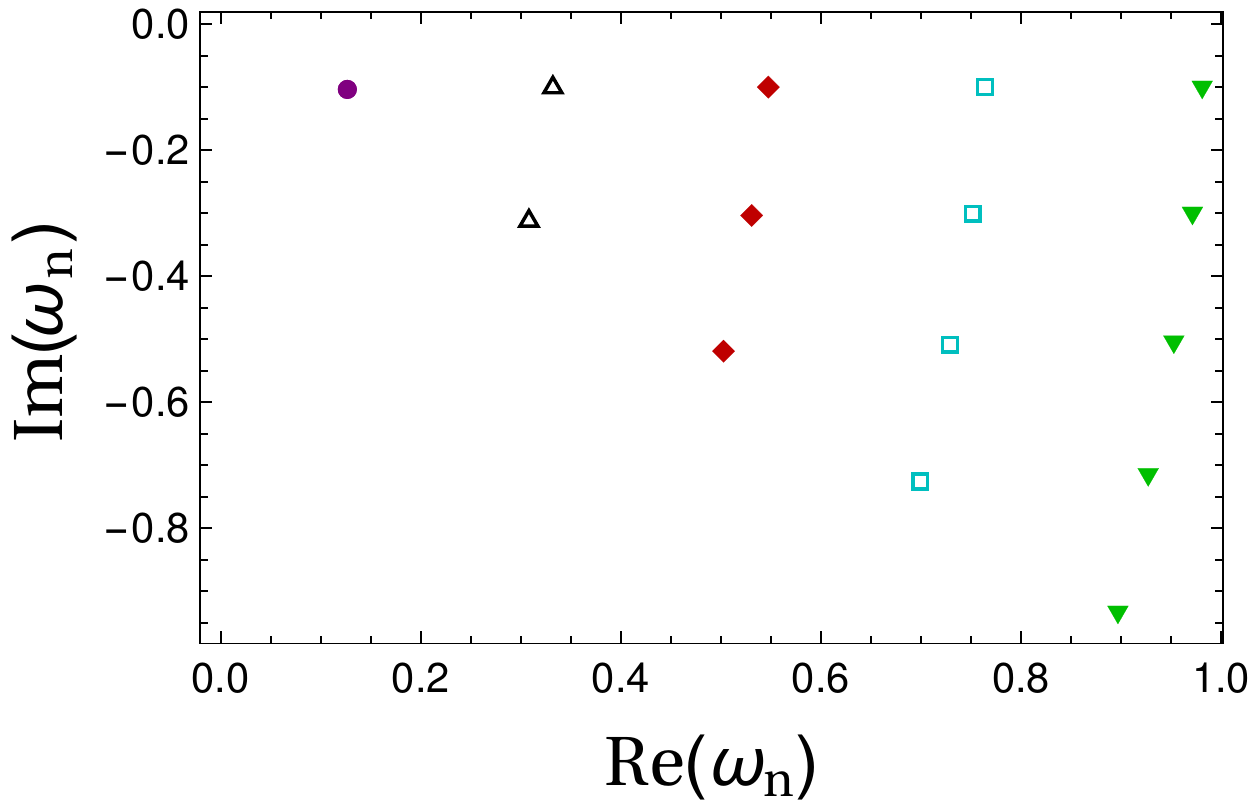}
\caption{
Quasinormal modes (scalar perturbations) $\omega_n \equiv \omega_R + i \omega_I$ for $M=1$. 
{\bf{LEFT:}} $\text{Im}(\omega_n)$ versus $\text{Re}(\omega_n)$ for $\alpha = 0$ in five different cases: 
i)   $l=0$ (purple point), 
ii)  $l=1$ (black triangle), 
iii) $l=2$ (red rhombus), 
iv)  $l=3$ (cyan square), and
 v)  $l=4$ (inverted green triangle)
{\bf{MIDDLE:}} Same as in the left panel, but for $\alpha = 0.1$. 
{\bf{RIGHT:}} Same as in the center panel, but for $\alpha = 0.2$. The first row 
corresponds to $Q_e=0.2$, the second to $Q_e=0.4$, the third to $Q_e=0.6$ and the last 
one to $Q_e=0.8$. Also, the first column of figures corresponds to tables I to IV, the second column corresponds to tables V to VIII, while the last column corresponds to tables IX to XII (see concrete values in each table).
}
\label{fig:frequencies_scalar}
\end{figure*}


\begin{table*} 
\centering
\caption{QN frequencies (scalar perturbations) for $M=1=\gamma, \alpha=0, Q_s=Q_e^2 \text{ and } Q_e=0.2$}
\resizebox{\columnwidth}{!}{%
\begin{tabular}{cccccc}
\hline
$n$ &  $l=0$ & $l=1$ & $l=2$ & $l=3$ & $l=4$\\
\hline
0    &  0.111249 -0.101024 i  &  0.294913 - 0.0979603 i  &  0.486919 -0.0969754 i   &  0.679931 -0.0967117 i  &  0.873273 -0.0966038 i \\
\hline
1    &                        &  0.26669 -0.306994 i    &  0.467277 -0.29621 i     &  0.665349 -0.292898 i   &  0.861754 -0.291498 i  \\
\hline
2    &                        &                         &  0.434076 -0.509528 i    &  0.63848 -0.496951 i    &  0.839802 -0.491312 i \\ 
\hline
3    &                        &                         &                          &  0.603595 -0.712548 i   &  0.809543 -0.698771 i \\
\hline
4    &                        &                         &                          &                         &  0.773858 -0.915726 i \\
\hline     
\end{tabular}
\label{table:First_set}
}
\end{table*}



\begin{table*} 
\centering
\caption{QN frequencies (scalar perturbations) for $M=1=\gamma, \alpha=0, Q_s=Q_e^2 \text{ and } Q_e=0.4$}
\resizebox{\columnwidth}{!}{%
\begin{tabular}{cccccc}
\hline
$n$ &  $l=0$ & $l=1$ & $l=2$ & $l=3$ & $l=4$\\
\hline
0    &   0.113615 -0.101748 i  &  0.301218 -0.0985422 i  &  0.497217 -0.0975905 i   &  0.694273 -0.0973332 i  &  0.891673 -0.0972279 i \\
\hline
1    &                         &  0.273688 -0.308355 i   &  0.478067 -0.297911 i    &  0.680055 -0.294689 i   &  0.880441 -0.293326 i  \\
\hline
2    &                         &                         &  0.445696, -0.511904 i   &  0.653861 -0.499691 i   & 0.859039 -0.494209 i \\ 
\hline
3    &                         &                         &                          &  0.619855 -0.715893 i   & 0.829539 -0.702517 i \\
\hline
4    &                         &                         &                          &                         &  0.794759 -0.920035 i \\
\hline     
\end{tabular}
\label{table:Second_set}
}
\end{table*}



\begin{table*} 
\centering
\caption{QN frequencies (scalar perturbations) for $M=1=\gamma, \alpha=0, Q_s=Q_e^2 \text{ and } Q_e=0.6$}
\resizebox{\columnwidth}{!}{%
\begin{tabular}{cccccc}
\hline
$n$ &  $l=0$ & $l=1$ & $l=2$ & $l=3$ & $l=4$\\
\hline
0    &   0.118615 -0.10242 i  &  0.312834 -0.0994461 i  &  0.516155 -0.0985552 i   &  0.720633 -0.0983114 i  &  0.925484 -0.0982119 i \\
\hline
1    &                         &  0.286614 -0.310356 i   &  0.497939 -0.30053 i    &  0.707105 -0.297483 i   &  0.914796 -0.296194 i  \\
\hline
2    &                         &                         &  0.467137 -0.515405 i   &  0.682184 -0.503884 i   & 0.894433 -0.498704 i \\ 
\hline
3    &                         &                         &                         &  0.649829 -0.720847 i   & 0.866368 -0.708226 i \\
\hline
4    &                         &                         &                         &                         &  0.833282 -0.926424 i \\
\hline     
\end{tabular}
\label{table:Third_set}
}
\end{table*}



\begin{table*} 
\centering
\caption{QN frequencies (scalar perturbations) for $M=1=\gamma, \alpha=0, Q_s=Q_e^2 \text{ and } Q_e=0.8$}
\resizebox{\columnwidth}{!}{%
\begin{tabular}{cccccc}
\hline
$n$ &  $l=0$ & $l=1$ & $l=2$ & $l=3$ & $l=4$\\
\hline
0    &  0.129042 -0.101423 i  &  0.332021 -0.100473 i  &  0.547375 -0.0996847 i  &  0.764056 -0.0994681 i  &  0.981166 -0.0993804 i \\
\hline
1    &                        &  0.307969 -0.312315 i  &  0.530756 -0.303444 i   &  0.751706 -0.300711 i  &  0.971406 -0.299552 i  \\
\hline
2    &                        &                        &  0.50261 -0.518801 i    &  0.728945 -0.508475 i  & 0.95281 -0.503815 i \\ 
\hline
3    &                        &                        &                         &  0.699358 -0.725739 i  & 0.927165 -0.714395 i \\
\hline
4    &                        &                        &                         &                        & 0.89691 -0.932771 i \\
\hline     
\end{tabular}
\label{table:Fourth_set}
}
\end{table*}



\begin{table*} 
\centering
\caption{QN frequencies (scalar perturbations) for $M=1=\gamma, \alpha=0.1, Q_e=0.2 \text{ and } Q_s=0.04$,}
\resizebox{\columnwidth}{!}{%
\begin{tabular}{cccccc}
\hline
$n$ &  $l=0$ & $l=1$ & $l=2$ & $l=3$ & $l=4$\\
\hline
0    &  0.111261 -0.101013 i  &  0.294914 -0.0979604 i &  0.486919 -0.0969754 i  &  0.679931 -0.0967118 i  &  0.873273 -0.0966038 i \\
\hline
1    &                        &  0.26669 -0.306995 i   &  0.467278 -0.29621 i    &  0.66535 -0.292898 i    &  0.861755 -0.291499 i  \\
\hline
2    &                        &                        &  0.434077 -0.509527 i   &  0.638481 -0.496952 i   & 0.839803 -0.491312 i \\ 
\hline
3    &                        &                        &                         &  0.603595 -0.712548 i   & 0.809543 -0.698771 i \\
\hline
4    &                        &                        &                         &                         &  0.77386 -0.915724 i \\
\hline     
\end{tabular}
\label{table:Fifth_set}
}
\end{table*}



\begin{table*} 
\centering
\caption{QN frequencies (scalar perturbations) for $M=1=\gamma, \alpha=0.1, Q_e=0.4 \text{ and } Q_s=0.159995$}
\resizebox{\columnwidth}{!}{%
\begin{tabular}{cccccc}
\hline
$n$ &  $l=0$ & $l=1$ & $l=2$ & $l=3$ & $l=4$\\
\hline
0    &  0.113734 -0.10164 i  &  0.301219 -0.0985427 i  &  0.497219 -0.0975908 i  &  0.694275 -0.0973335 i  &  0.891675 -0.0972282 i \\
\hline
1    &                       &  0.273686 -0.30836 i   &  0.478069 -0.297912 i    &  0.680057 -0.29469 i   &  0.880444 -0.293327 i  \\
\hline
2    &                       &                        &  0.445699 -0.511904 i   &  0.653863 -0.499692 i  & 0.859041 -0.494211 i \\ 
\hline
3    &                       &                        &                         &  0.619857 -0.715895 i  & 0.829542 -0.702519 i \\
\hline
4    &                       &                        &                         &                         & 0.794762 -0.920037 i \\
\hline     
\end{tabular}
\label{table:Sixth_set}
}
\end{table*}



\begin{table*} 
\centering
\caption{QN frequencies (scalar perturbations) for $M=1=\gamma, \alpha=0.1, Q_e=0.6 \text{ and } Q_s=0.35986$}
\resizebox{\columnwidth}{!}{%
\begin{tabular}{cccccc}
\hline
$n$ &  $l=0$ & $l=1$ & $l=2$ & $l=3$ & $l=4$\\
\hline
0    &  0.118441 -0.102568 i  &  0.312834 -0.0994459 i  &  0.516155 -0.0985551 i   &  0.720632 -0.0983113 i  &  0.925483 -0.0982118 i \\
\hline
1    &                        &  0.286617 -0.310352 i   &  0.497939 -0.300529 i    &  0.707105 -0.297483 i   &  0.914795 -0.296193 i  \\
\hline
2    &                        &                         &  0.467137 -0.515405 i    &  0.682184 -0.503883 i  & 0.894432 -0.498703 i \\ 
\hline
3    &                        &                         &                          &  0.64983 -0.720845 i   & 0.866366 -0.708226 i \\
\hline
4    &                        &                         &                          &                         &  0.833278 -0.926428 i \\
\hline     
\end{tabular}
\label{table:Seventh_set}
}
\end{table*}



\begin{table*} 
\centering
\caption{QN frequencies (scalar perturbations) for $M=1=\gamma, \alpha=0.1, Q_e=0.8 \text{ and } Q_s=0.638614$}
\resizebox{\columnwidth}{!}{%
\begin{tabular}{cccccc}
\hline
$n$ &  $l=0$ & $l=1$ & $l=2$ & $l=3$ & $l=4$\\
\hline
0    &  0.126487 -0.10344 i  &  0.332038 -0.100468 i  &  0.547379 -0.0996838 i  &  0.764062 -0.0994672 i  &  0.981175 -0.0993795 i \\
\hline
1    &                       &  0.308088 -0.312194 i  &  0.530757 -0.303443 i   &  0.751713 -0.300708 i   &  0.971415 -0.299549 i  \\
\hline
2    &                       &                        &  0.502601 -0.51881 i    &  0.728954 -0.508468 i   & 0.952819 -0.50381 i \\ 
\hline
3    &                       &                        &                         &  0.699371 -0.725727 i   & 0.927173 -0.714389 i \\
\hline
4    &                       &                        &                         &                         &  0.896915 -0.932767 i \\
\hline     
\end{tabular}
\label{table:Eighth_set}
}
\end{table*}



\begin{table*} 
\centering
\caption{QN frequencies (scalar perturbations) for $M=1=\gamma, \alpha=0.2, Q_e=0.2 \text{ and } Q_s=0.04$}
\resizebox{\columnwidth}{!}{%
\begin{tabular}{cccccc}
\hline
$n$ &  $l=0$ & $l=1$ & $l=2$ & $l=3$ & $l=4$\\
\hline
0    &  0.111263 -0.101011 i  &  0.294914 -0.0979604 i  &  0.486919 -0.0969755 i  &  0.679932 -0.0967118 i  &  0.873274 -0.0966039 i \\
\hline
1    &                        &  0.26669 -0.306995 i    &  0.467278 -0.29621 i    &  0.66535 -0.292898 i   &  0.861755 -0.291499 i  \\
\hline
2    &                        &                         &  0.434077 -0.509528 i    &  0.638481 -0.496952 i  & 0.839803 -0.491313 i \\ 
\hline
3    &                        &                         &                          &  0.603596 -0.712548 i   & 0.809544 -0.698771 i \\
\hline
4    &                        &                         &                          &                         &  0.773859 -0.915726 i \\
\hline     
\end{tabular}
\label{table:Ninth_set}
}
\end{table*}



\begin{table*} 
\centering
\caption{QN frequencies (scalar perturbations) for $M=1=\gamma, \alpha=0.2, Q_e=0.4 \text{ and } Q_s=0.159989$}
\resizebox{\columnwidth}{!}{%
\begin{tabular}{cccccc}
\hline
$n$ &  $l=0$ & $l=1$ & $l=2$ & $l=3$ & $l=4$\\
\hline
0    &  0.113747 -0.101628 i  &  0.301218 -0.0985423 i  &  0.497217 -0.0975905 i  &  0.694273 -0.0973331 i  &  0.891672 -0.0972278 i \\
\hline
1    &                        &  0.273685 -0.308359 i   &  0.478067 -0.297911 i   &  0.680055 -0.294689 i   &  0.880441 -0.293326 i  \\
\hline
2    &                        &                         &  0.445697 -0.511903 i   &  0.65386 -0.499691 i    & 0.859038  -0.494209 i \\ 
\hline
3    &                        &                         &                         &  0.619855 -0.715892 i   & 0.829539 -0.702517 i \\
\hline
4    &                        &                         &                         &                         &  0.794759 -0.920034 i \\
\hline     
\end{tabular}
\label{table:Tenth_set}
}
\end{table*}



\begin{table*} 
\centering
\caption{QN frequencies (scalar perturbations) for $M=1=\gamma, \alpha=0.2, Q_e=0.6 \text{ and } Q_s=0.359721$}
\resizebox{\columnwidth}{!}{%
\begin{tabular}{cccccc}
\hline
$n$ &  $l=0$ & $l=1$ & $l=2$ & $l=3$ & $l=4$\\
\hline
0    &  0.118456 -0.102554 i  &  0.312835 -0.0994459 i  &  0.516156 -0.0985552 i  &  0.720633 -0.0983114 i  &  0.925484 -0.0982119 i \\
\hline
1    &                        &  0.286619 -0.310351 i   &  0.497939 -0.30053 i    &  0.707106 -0.297483 i   &  0.914797 -0.296193 i  \\
\hline
2    &                        &                         &  0.467136 -0.515406 i   &  0.682185 -0.503884 i  & 0.894434 -0.498704 i \\ 
\hline
3    &                        &                         &                         &  0.649832 -0.720845 i   & 0.866368 -0.708226 i \\
\hline
4    &                        &                         &                         &                         &  0.833281 -0.926427 i \\
\hline     
\end{tabular}
\label{table:Eleventh_set}
}
\end{table*}



\begin{table*} 
\centering
\caption{QN frequencies (scalar perturbations) for $M=1=\gamma, \alpha=0.2, Q_e=0.8 \text{ and } Q_s=0.637252$}
\resizebox{\columnwidth}{!}{%
\begin{tabular}{cccccc}
\hline
$n$ &  $l=0$ & $l=1$ & $l=2$ & $l=3$ & $l=4$\\
\hline
0    &  0.126564 -0.103373 i  &  0.332038 -0.100467 i  &  0.547383 -0.099683 i  &  0.764069 -0.0994664 i  &  0.981184 -0.0993787 i \\
\hline
1    &                        &  0.30808 -0.312201 i   &  0.530762 -0.30344 i   &  0.751721 -0.300705 i   &  0.971424 -0.299547 i  \\
\hline
2    &                        &                        &  0.502607 -0.518804 i  &  0.728962 -0.508463 i  & 0.952829 -0.503806 i \\ 
\hline
3    &                        &                        &                        &  0.699382 -0.725717 i   & 0.927184 -0.714382 i \\
\hline
4    &                        &                        &                        &                         &  0.896927 -0.932758 i \\
\hline     
\end{tabular}
\label{table:Twelfth_set}
}
\end{table*}


Our results are summarized in Tables I-XII, and for better visualization are shown in Fig.~\ref{fig:frequencies_scalar}. The real part of the modes decreases with $n$ and increases with $\alpha,l,Q_e$, whereas the absolute value of the imaginary part of the frequencies increases with $n,Q_e$ and decreases with $\alpha,l$. Our results are similar to the ones of \cite{Chowdhury:2018izv} for a massless and electrically neutral scalar field.

\smallskip

The QN modes for neutral scalar perturbations of black holes in the three-dimensional EMD theory \cite{Chan:1994qa} were computed in \cite{Fernando:2003ai}, where an exact analytical expression for the spectrum was obtained, and all modes were found to be purely imaginary. On the contrary, in the parameter space considered here, all modes are found to have a non-vanishing real part.

\section{Conclusions}

To summarize, in the present work we have computed the quasinormal spectrum for scalar perturbations of a black hole with a scalar hair in the Einstein-Maxwell-dilaton model in four space-time dimensions assuming a non-trivial scalar potential for the dilaton. The black hole is characterized by an electric charge as well as a scalar charge, which nevertheless depends on the other properties of the black hole solution. After summarizing the model, the field equations and the solution, we perturbed the black hole with a test massless scalar field, investigating its propagation into a fixed gravitational background. The effective potential barrier of the corresponding Schr{\"o}dinger-like equation for scalar perturbations was obtained. Finally, the quasinormal frequencies were computed adopting the semi-analytic WKB method of 6th order. Contrary to the three-dimensional EMD theory, where the QN frequencies for scalar perturbations of black holes were found to be purely imaginary, here all modes have a non-vanishing real part. Moreover, all modes were found to be stable. Our results have been shown in tables as well as in in figures, for better visualization. The impact on the spectrum of the overtone number, the angular degree, the charges of the black hole as well as the magnitude of the dilaton potential has been investigated in detail.



\section*{Acknowlegements}

We wish to thank the anonymous reviewers for useful comments and suggestions as well as for pointing out an error in our original computation. The author \'A.~R. acknowledges DI-VRIEA for financial support through Proyecto Postdoctorado 2019 VRIEA-PUCV. The author G.~P. thanks the Fun\-da\c c\~ao para a Ci\^encia e Tecnologia (FCT), Portugal, for the financial support to the Center for Astrophysics and Gravitation-CENTRA, Instituto Superior T\'ecnico, Universidade de Lisboa, through the Project No.~UIDB/00099/2020.


\bibliographystyle{unsrt} 
\bibliography{Bibliography,Bibliography_New}  

\end{document}